\documentclass[12pt]{article}
\pdfoutput=1

\usepackage{subcaption}
\usepackage[dvipsnames]{xcolor}
\usepackage[totalheight = 23cm, totalwidth = 17cm]{geometry}
\usepackage{amssymb,amsmath,amsfonts,amsbsy,graphicx}
\usepackage{bm,color}
\usepackage{cite}
\usepackage[%
             colorlinks=true,urlcolor=MidnightBlue,linkcolor=MidnightBlue,citecolor=OliveGreen,
             pdfpagelabels=true,hypertexnames=true,
            plainpages=false,naturalnames=false,
             ]{hyperref}

\newcommand{\mpl}{m_{\rm Pl}}
\newcommand{\fnl}{f_{\rm NL}}
\newcommand{\calP}{{\cal P}}
\newcommand{\calR}{{\cal R}}

\begin{document}

\begin{titlepage}

\rightline{\footnotesize{APCTP-Pre2017-002}}

\begin{center}

\vskip 1.0 cm

{\Large \bf Shapes and features of the primordial bispectrum}

\vskip 1.0cm

{\large
Jinn-Ouk Gong$^{a,b}$, Gonzalo A. Palma$^{c}$ and Spyros Sypsas$^{c}$
}

\vskip 0.5cm

{\it
$^{a}$Asia Pacific Center for Theoretical Physics, Pohang 37673, Korea
\\
$^{b}$Department of Physics, Postech, Pohang 37673, Korea
\\
$^{c}$Departamento de F\'isica, FCFM, Universidad de Chile, Santiago 837.0415, Chile
}

\vskip 1.2cm

\end{center}

\begin{abstract}

If time-dependent disruptions from slow-roll occur during inflation, the correlation functions of the primordial curvature perturbation should have scale-dependent features, a case which is marginally supported from the cosmic microwave background (CMB) data. We offer a new approach to analyze the appearance of such features in the primordial bispectrum that yields new consistency relations and justifies the search of oscillating patterns modulated by orthogonal and local templates. Under the assumption of sharp features, we find that the cubic couplings of the curvature perturbation can be expressed in terms of the bispectrum in two specific momentum configurations, for example local and equilateral. This allows us to derive consistency relations among different bispectrum shapes, which in principle could be tested in future CMB surveys. Furthermore, based on the form of the consistency relations, we construct new two-parameter templates for features that include all the known shapes.

\end{abstract}

\end{titlepage}

\newpage
\setcounter{page}{1}

\section{Introduction}

Inflation~\cite{inflation} is the leading idea providing a satisfactory description of the evolution of the very early universe before the standard hot big bang commences. Some of its predictions, like homogeneity, isotropy and flatness of the observable patch of the universe, have been verified with unprecedented precision, with the most up-to-date observational surveys pointing to a nearly scale-invariant, Gaussian distribution of the primordial perturbations~\cite{Ade:2015xua}. However, the theory into which the inflationary mechanism is embedded is still elusive and there remains a number of questions regarding inflation model building~\cite{inflation-model}. String theory provides such an ultraviolet (UV) complete framework, a general characteristic of which is the presence of a variety of fields and peculiarities which can alter the otherwise smooth dynamics of inflation.

On the observational side, the Planck mission on the cosmic microwave background (CMB) has provided data of unprecedented precision on the CMB power spectrum~\cite{Aghanim:2015xee}, which supports most inflationary predictions~\cite{Ade:2015lrj}. There seems however to exist several anomalies in the power spectrum, including deviations from scale-invariance indicated by outliers beyond 2$\sigma$ significance at several multipole ranges~\cite{dev-scaleinv}. The appearance of these features has sparked intensive efforts to study such a possible scale-dependence~\cite{featuresearch} which is indeed predicted by such embeddings of inflation in UV complete frameworks -- see e.g.~\cite{strings}.

Observable features of the power spectrum can be produced in various ways: these include for example  heavy degrees of freedom, resulting in a non-trivial speed of sound~\cite{cs-heavy}, and features in the inflaton potential~\cite{Starobinsky:1992ts,inflatonpotential}, resulting in sudden variations of the slow-roll parameters. These mechanisms provide deviations from the standard single-field canonical slow-roll inflation. Furthermore, it is expected that such a behaviour will also affect the bispectrum (as well as all the higher-order spectra), whose shape as a function of the three momenta depends strongly on the class of models under consideration. For example, non-canonical inflation leads to equilateral or orthogonal non-Gaussianity and multi-field inflation leads to a local shape~\cite{ngreview}. This implies that joint analysis of the power spectrum and bispectrum would lead to stronger constraints~\cite{generalizedfeature,jointanalysis}.

Indeed, a possible detection of features in the bispectrum would give us deeper insights into the details of inflation, since on one hand it would increase the statistical significance of these anomalies while on the other hand it would shed light on the mechanism responsible for them. However, observational data on the primordial bispectrum are still quite poor, a fact that highly limits the ability of observational surveys to guide theoretical searches towards the correct answer. Hence, what can be done at this stage is to construct theoretically motivated bispectrum templates, which can then be searched for in the data, and consistency relations or tests of certain assumptions which can be used in sparse data sets.

In this article, we construct such consistency relations for the primordial bispectrum in the general case where sudden deviations from slow-roll occur in the potential and/or the kinetic parameters, like the Hubble parameter or the speed of sound of the curvature perturbation. Therefore, if features exist in the correlation functions then we can predict patterns that the bispectrum should follow. Furthermore, motivated from our relations, we show that it is possible for the orthogonal and local templates to appear with an oscillatory profile, typical of inflationary features, complementing the current use of templates modulated by equilateral and flattened shapes~\cite{Ade:2015ava}. Our analysis is also suitable for the case of multi-frequency oscillating patterns in the three-point function~\cite{Ade:2015ava}.

Specifically, we derive a set of shape consistency relations, \eqref{eq:con-rel1} and \eqref{eq:con-rel3} in the sense that measurements of the bispectrum at any three sets of modes with momenta $(k_1,k_2,k_3)_i$ for $i=1,2,3$ should be related. We expect that such consistency tests will be of use in prospect of the large-scale structure surveys, which are expected to provide information on the bispectrum at multiple $k$-modes, greatly improving the constraints on features~\cite{lss-feature}. Then, based on the qualitative form of these consistency relations, we also propose two bispectrum templates for features that can include all the known modulating shapes such as the equilateral, flattened, orthogonal and local ones.

The article is organised as follows: in Section~\ref{sec:bispectrum} we present the general form of the bispectrum in the presence of deviations from usual slow-roll evolution during inflation. In Section~\ref{sec:shapeconsistency} we derive the bispectrum consistency relations, while in Section~\ref{sec:template} we study the component shapes that are involved in the consistency relations and compare them with the already used templates. Finally, we conclude in Section~\ref{sec:conclusion}.

\section{Bispectrum with sharp features}
\label{sec:bispectrum}

The cubic action for the curvature perturbation can be, after integration by parts, written as~\cite{Burrage:2011hd}
\begin{align}
\label{eq:S3-full}
S_3 = \int d^4x a^3\epsilon\mpl^2 
& \bigg[ \frac{\epsilon - \eta -3(1-c_s^2)}{c_s^4}\dot{\calR}^2\calR + \frac{\epsilon + \eta + 1-c_s^2 -2s}{a^2c_s^2} \calR(\nabla\calR)^2 
\nonumber\\ 
& + \frac{1}{H}\left( \frac{1-c_s^2}{c_s^4} - \frac{2\lambda}{\epsilon H^2} \right) \dot{\calR}^3 
+ \frac{1}{4a^4}(\nabla\chi)^2 \nabla^2 \calR + \frac{\epsilon-4}{2\epsilon a^4} \nabla^2\chi \partial_i \calR \partial^i\chi +\frac{f(\calR)}{\epsilon a^3} \frac{\delta S_{2}}{\delta \calR} \bigg] \, ,
\end{align}
up to boundary terms. Here $\chi$ is determined by the constraint $\nabla^2\chi = a^2 \epsilon \dot{\calR}/c_s^2$ and $\epsilon \equiv -\dot{H}/H^2$, $\eta \equiv \dot{\epsilon}/(H\epsilon)$ and $s \equiv \dot{c_s}/(Hc_s)$. $\lambda$ is a model-dependent parameter~\cite{lambdaparameter} and the last term multiplied by the linear equation of motion $\delta{S}_2/\delta\calR$ can be removed by a field redefinition~\cite{Maldacena:2002vr}.

In what follows we will study the case where background parameters experience temporary deviations from slow-roll. In general, there are two extreme cases that could be studied separately: ({\it i}) features leading to bumps or oscillatory profiles in the CMB power spectrum in certain multipoles with a characteristic width $\Delta k$ such that $\Delta k / k \ll 1$ and ({\it ii}) deviations leading to smooth changes of the overall spectrum, in the form of running, with a characteristic width $\Delta k / k \gg 1$. Here we will focus on the first case, and for such an effect to be observable in the CMB, which covers a few $e$-folds during inflation, such a variation should indeed happen in a few $e$-folds. However, this is in no way unambiguously implied by the current data -- but it is a reasonable approximation that captures the essence of the problem.

Having stressed that, to quantify this statement, we allow any background parameter like $b=\{\epsilon,c_s\}$ satisfy
\begin{equation} 
\frac{|\dot{b}|}{H} \gg |b| \, ,
\end{equation}
which can be more conveniently written in terms of the conformal time $\tau$ as
\begin{equation} \label{sharp-cond}
\tau \left|b'\right| \gg |b| \, .
\end{equation}
Such a condition drastically simplifies the form of the full cubic action \eqref{eq:S3-full} to~\cite{Mooij:2015cxa}
\begin{equation}
\label{eq:S3}
S_3 \approx \int d\tau d^3x a^2\epsilon\mpl^2 
\left[ c_1\calR'^2\calR + c_2 \calR(\nabla\calR)^2 \right] \, ,
\end{equation}
which is the form that we will use. The coefficients $c_i$ contain the variations of $\epsilon$ and $c_s$~\cite{Palma:2014hra}. Thus given that their amplitudes are constrained to remain small from various observations, their rapid variations can give rise to sharp features. Note that in~\cite{Mooij:2015cxa}, a reasonable but ad-hoc relation between $\eta$ and $s$ was imposed, so that the two dimensionless coefficients $c_1$ and $c_2$ in \eqref{eq:S3}, which are in principle independent parameters in the effective field theory point of view~\cite{Cheung:2007st}, were related by a single parameter. Here we do not use any such relation and we keep $c_i$ free.

With the leading de Sitter mode function solution for $\calR$ with $c_s=1$,
\begin{equation}
\widehat\calR_k(\tau) = \frac{iH}{\sqrt{4\epsilon k^3}\mpl} (1+ik\tau ) e^{-ik\tau} \, ,
\end{equation}
we can follow the standard in-in formalism to compute the bispectrum as
\begin{align}
\label{eq:bi}
B_\calR(k_1,k_2,k_3) = \frac{(2\pi)^4\calP_\calR^2}{(k_1k_2k_3)^3} 2 \Re & 
\left\{ \int_{-\infty}^0 d\tau \frac{c_1}{8} \left[ (k_1k_2)^2 (1-ik_3\tau) + \text{2 perm} \right]  ie^{iK\tau} \right.
\nonumber\\
& \left. + \int_{-\infty}^0 \frac{d\tau}{\tau^2} c_2 \frac{k_1^2+k_2^2+k_3^2}{16} 
(1-ik_1\tau) (1-ik_2\tau) (1-ik_3\tau) ie^{iK\tau} \right\} \, ,
\end{align}
where
\begin{equation}
\calP_\calR = \frac{H^2}{8\pi^2\epsilon\mpl^2}
\end{equation}
is the featureless flat power spectrum and $K \equiv k_1+k_2+k_3$. By extending $c_i$'s oddly in the conformal time, i.e. $c_i(-\tau) = -c_i(\tau)$ and replacing one $k$ in the integrands with one derivative with respect to $\tau$ acting on $e^{iK\tau}$, we can rewrite \eqref{eq:bi} as
\begin{align}
\label{eq:bi2}
S_\calR(k,x,y) = 
\int_{-\infty}^\infty d\tau \frac{i}{8} \frac{e^{i(1+x+y)k\tau}}{xyk} 
& \left[ -\frac{xyS_1(x,y)}{1+x+y}(c_1\tau)''' + \frac{xyS_2(x,y)}{1+x+y}(c_2\tau)''' \right.
\nonumber\\
& \left.  -\frac{F_1(x,y)}{1+x+y}c_1'' - \frac{F_2(x,y)}{1+x+y}c_2'' + \left( 1+x^2+y^2 \right) \frac{c_2'}{\tau}
 \right]  \, ,
\end{align}
where we have defined the dimensionless shape function $S_\calR(k_1,k_2,k_3)$ as
\begin{equation}
B_\calR(k_1,k_2,k_3) \equiv \frac{(2\pi)^4\calP_\calR^2}{(k_1k_2k_3)^2} S_\calR(k_1,k_2,k_3) \, ,
\end{equation}
and the partial shapes
\begin{equation} \label{s-f-defs}
\begin{split}
& S_1(x,y )= \frac{x+y+xy}{(1+x+y)^2} \, , \quad S_2(x,y) = \frac{1+x^2+y^2}{2(1+x+y)^2} \, , 
\\
& F_1(x,y) = \frac{x^2+y^2+x^2y^2}{1+x+y} \, , \quad F_2(x,y) = \frac{(1+x^2+y^2)(x+y+xy)}{2(1+x+y)} \, ,
\end{split}
\end{equation}
with
\begin{equation}
k_1 \equiv k \, , \quad k_2 \equiv xk \, , \quad k_3 \equiv yk \, .
\end{equation}
Note that being dimensionless, we use the arguments of the shape function $S_\calR$ as $(k_1,k_2,k_3)$ and $(k,x,y)$ interchangeably.

\section{Shape consistency relations}
\label{sec:shapeconsistency}

\subsection{Bispectrum in terms of two general momentum configurations}

In the previous section, we have obtained the general shape function $S_\calR(k,x,y)$ without any approximation for the integrands. We now make further use of the sharp feature condition \eqref{sharp-cond}. It follows that for any $c_i$, higher time derivatives dominate over lower ones so that
\begin{equation}
\label{sharp-der}
(c_i\tau)''' = c_i'''\tau + 3c_i'' \approx c_i'''\tau \, ,
\end{equation}
allowing us to keep only the $c_i'''\tau$ terms. Thus, under the sharp feature assumption, \eqref{eq:bi2} is simply written as
\begin{equation}
\label{eq:bi3}
S_\calR(k,x,y) = 
\int_{-\infty}^\infty d\tau \frac{i}{8} \frac{e^{i(1+x+y)k\tau}}{k} 
\left[ -\frac{S_1(x,y)}{1+x+y}c_1'''\tau + \frac{S_2(x,y)}{1+x+y}c_2'''\tau \right] \, .
\end{equation}
As we can see, once we are interested in a certain configuration fixed by specific $x$ and $y$, the featured bispectrum behaves according to the two model-dependent coefficients $c_1'''\tau$ and $c_2'''\tau$. This means that by fixing two triangle configurations $\Delta_1 = (x_1,y_1)$ and $\Delta_2 = (x_2,y_2)$ and solving for $c_1'''\tau$ and $c_2'''\tau$, the full bispectrum can be completely specified.

For this, following the method presented in~\cite{invert}, we first invert \eqref{eq:bi3} by multiplying both sides by $e^{iK\tau'}$ and integrating over $K$ to obtain
\begin{equation}
\label{eq:bi-sharp}
\frac{4}{i\pi} \int_{-\infty}^\infty dk ke^{-i(1+x+y)k\tau} S_\calR(k,x,y) 
\equiv {\cal S}^\Delta(\tau) 
= -\frac{S_1(x,y)}{(1+x+y)^2}c_1'''\tau + \frac{S_2(x,y)}{(1+x+y)^2}c_2'''\tau \, ,
\end{equation}
which gives us the time-dependent coefficients $c_i'''\tau$ in terms of the bispectrum. Indeed, we can write $c_1'''\tau$ and $c_2'''\tau$ in terms of two shapes as
\begin{equation}
\begin{split}
c_1'''\tau &= -(1+x_1+y_1)^2\alpha_1{\cal S}^{\Delta_1}(\tau) + (1+x_2+y_2)^2\beta_1{\cal S}^{\Delta_2}(\tau) \, ,
\\
c_2'''\tau &= -(1+x_1+y_1)^2\alpha_2{\cal S}^{\Delta_1}(\tau) + (1+x_2+y_2)^2\beta_2{\cal S}^{\Delta_2}(\tau) \, ,
\end{split}
\end{equation}
with
\begin{equation} 
\label{eq:ab-def}
\begin{split}
\alpha_1 & = \frac{S_2(x_2,y_2)}{S_1(x_1,y_1)S_2(x_2,y_2)-S_1(x_2,y_2)S_2(x_1,y_1)} \, , \quad
\beta_1  = \frac{S_2(x_1,y_1)}{S_1(x_1,y_1)S_2(x_2,y_2)-S_1(x_2,y_2)S_2(x_1,y_1)} \, , 
\\
\alpha_2 & = \frac{S_1(x_2,y_2)}{S_1(x_1,y_1)S_2(x_2,y_2)-S_1(x_2,y_2)S_2(x_1,y_1)} \, , \quad
\beta_2 = \frac{S_1(x_1,y_1)}{S_1(x_1,y_1)S_2(x_2,y_2)-S_1(x_2,y_2)S_2(x_1,y_1)}  \, .
\end{split}
\end{equation}
Plugging them back to \eqref{eq:bi3} and performing the integrals, we can write a general shape function $S_\calR(k,x,y)$ in terms of two specified configurations as
\begin{align}
\label{eq:con-rel1}
S_\calR(k,x,y) & = \Big[ \alpha_1 S_1(x,y) - \alpha_2 S_2(x,y) \Big] S_\calR \left( \frac{1+x+y}{1+x_1+y_1}k,x_1,y_1 \right) 
\nonumber\\
& \quad + \Big[ -\beta_1 S_1(x,y) + \beta_2 S_2(x,y) \Big] S_\calR \left( \frac{1+x+y}{1+x_2+y_2}k,x_2,y_2 \right) \, .
\end{align}
For example, by fixing $\Delta_1=(1,1)$ and $\Delta_2=(1/2,1/2)$ we obtain
\begin{align}
\label{eq:eq-flat}
S_\calR(k,x,y) & = \frac{18(x+y+xy) - 15(1+x^2+y^2)}{(1+x+y)^2} S_\calR^{} \left(\frac{1+x+y}{3}k,1,1\right)
\nonumber\\
& \quad + \frac{-16(x+y+xy) + 16(1+x^2+y^2)}{(1+x+y)^2} S_\calR^{} \left(\frac{1+x+y}{2}k,\frac{1}{2},\frac{1}{2}\right) \, ,
\end{align}
implying that the amplitude of the bispectrum at any point $(k,x,y)$ should be related to the corresponding values at $\left(\frac{1+x+y}{3}k,1,1\right)$ and $\left(\frac{1+x+y}{2}k,1/2,1/2\right)$.

\subsection{Bispectrum including the squeezed configuration}

Including the squeezed configuration of the bispectrum, say, $x_{sq} \approx 1$ and $y_{sq} \to 0$, needs more care. This is because the sub-leading $F_1(x,y)$ and $F_2(x,y)$ terms in \eqref{eq:bi2} are boosted by $1/y_{sq}$ and become the leading contributions to the bispectrum. Therefore, we need to keep both the leading and sub-leading terms in the sharp feature assumption \eqref{sharp-cond}, so we begin with
\begin{align}
\label{eq:shape-sq}
S_\calR(k,x,y) = \int_{-\infty}^\infty d\tau \frac{i}{8} \frac{e^{i(1+x+y)k\tau}}{k} 
& \left[ \frac{-S_1(x,y)}{1+x+y}c_1'''\tau + \frac{S_2(x,y)}{1+x+y}c_2'''\tau \right.
\nonumber\\
& \left. - \frac{F_1(x,y)}{xy(1+x+y)}c_1'' - \frac{F_2(x,y)}{xy(1+x+y)}c_2'' \right] \, .
\end{align}
We first apply the squeezed configuration. Then $F_i$ terms are boosted by $1/y_{sq}$ so become dominant over $S_i$ terms. Since $F_1(x\to1,y\to0)=F_2(x\to1,y\to0)=1/2$, we can write
\begin{equation}
\label{eq:sq-inv0}
k x_{sq}y_{sq} S_\calR(k,x_{sq},y_{sq}) = \frac{1}{32i} 
\int_{-\infty}^\infty d\tau e^{2ik\tau} \left( c_1''+c_2'' \right) \, .
\end{equation}
Next, considering another configuration away from the squeezed one $(x_2,y_2)$, $S_i$ terms are dominant based on our assumption of sharp feature. That is,
\begin{equation}
\label{eq:shape2-sq}
S_\calR(k,x_2,y_2) = \int_{-\infty}^\infty d\tau \frac{i}{8} \frac{e^{i(1+x_2+y_2)k\tau}}{k} 
\left[ \frac{-S_1(x_2,y_2)}{1+x_2+y_2}c_1'''\tau + \frac{S_2(x_2,y_2)}{1+x_2+y_2}c_2'''\tau \right] \, .
\end{equation}
Replacing $\tau$ with a derivative with respect to $k$ acting on the exponential factor, the right hand side of \eqref{eq:shape2-sq} becomes a total derivative with respect to $k$ as
\begin{equation}
S_\calR(k,x_2,y_2) = \frac{1}{8k} \frac{d}{dk} \int_{-\infty}^\infty d\tau \frac{e^{i(1+x_2+y_2)k\tau}}{(1+x_2+y_2)^2} 
\Big[ -S_1(x_2,y_2)c_1''' + S_2(x_2,y_2)c_2''' \Big] \, .
\end{equation}
Multiplying both sides with $k$ and integrating over it, we have 
\begin{align}
\label{eq:x2y2-inv0}
\Sigma_2(k,x_2,y_2) \equiv \int_0^k dq qS_\calR(q,x_2,y_2) & = \frac{1}{8} \int_{-\infty}^\infty d\tau \frac{e^{i(1+x_2+y_2)k\tau}}{(1+x_2+y_2)^2} \Big[ -S_1(x_2,y_2)c_1''' + S_2(x_2,y_2)c_2''' \Big]
\nonumber\\
& = \frac{1}{8} \int_{-\infty}^\infty d\tau ik \frac{e^{i(1+x_2+y_2)k\tau}}{1+x_2+y_2} \Big[ S_1(x_2,y_2)c_1'' - S_2(x_2,y_2)c_2'' \Big] \, ,
\end{align}
where for the second equality we have integrated by parts over $c_i'''$ to directly combine with \eqref{eq:sq-inv0}. Thus, we can invert \eqref{eq:sq-inv0} and \eqref{eq:x2y2-inv0} to obtain respectively
\begin{equation}
\begin{split}
c_1''+c_2'' & = \frac{32i}{\pi} x_{sq}y_{sq} \int_{-\infty}^\infty dk ke^{-2ik\tau} S_\calR(k,x_{sq},y_{sq}) \, .
\\
\frac{-S_1(x_2,y_2)}{(1+x_2+y_2)^2}c_1'' + \frac{S_2(x_2,y_2)}{(1+x_2+y_2)^2}c_2'' & = \frac{4i}{\pi} \int_{-\infty}^\infty \frac{dk}{k} e^{-i(1+x_2+y_2)k\tau} \Sigma_2(k,x_2,y_2)  \, .
\end{split}
\end{equation}
From this algebraic system it is trivial to find the solutions for $c_1''$ and $c_2''$. By differentiating them and then manipulating the momentum integral accordingly, we can easily find $c_i'''\tau$.

Plugging $c_i''$ and $c_i'''\tau$ back into \eqref{eq:shape-sq} we may write the general shape function in terms of the squeezed configuration and another one as
\begin{align}
\label{eq:con-rel3}
S_\calR(k,x,y) & = \left[ -\gamma_1S_1(x,y) + \gamma_2S_2(x,y) + \frac{\gamma_1}{2}\frac{F_1(x,y)}{xy} + \frac{\gamma_2}{2}\frac{F_2(x,y)}{xy} \right] x_{sq}y_{sq} S_\calR\left( \frac{1+x+y}{2}k,x_{sq},y_{sq} \right)
\nonumber\\
& \quad + \frac{1}{2} \Big[ -\gamma_1S_1(x,y) + \gamma_2S_2(x,y) \Big] x_{sq}y_{sq} \frac{\partial S_\calR(p,x_{sq},y_{sq})}{\partial\log{p}} \bigg|_{p=(1+x+y)k/2}
\nonumber\\
& \quad + \frac{\gamma_3}{(1+x+y)^2} \left[ -\frac{F_1(x,y)}{xy} + \frac{F_2(x,y)}{xy} \right] \frac{1}{k^2} \Sigma_2 \left( \frac{1+x+y}{1+x_2+y_2}k,x_2,y_2 \right)
\nonumber\\
& \quad +  S_\calR\left( \frac{1+x+y}{1+x_2+y_2}k,x_2,y_2 \right) \, ,
\end{align}
where
\begin{equation} 
\label{g-defs}
\gamma_1 = 8S_2(x_2,y_2) \, , 
\quad 
\gamma_2 = 8S_1(x_2,y_2) \, , 
\quad
\gamma_3 = 2(1+x_2+y_2)^2 \, .
\end{equation}
An application of \eqref{eq:con-rel3} is, for example, a relation between the bispectrum in the squeezed, equilateral and enfolded configurations:
\begin{align}
\label{eq:sq-eq-enf}
S_\calR(k,1,1) & = \frac{61}{24} x_{sq}y_{sq} S_\calR\left( \frac{3}{2}k,x_{sq},y_{sq} \right)
- \frac{1}{24} x_{sq}y_{sq} \frac{\partial S_\calR(p,x_{sq},y_{sq})}{\partial\log{p}} \bigg|_{p=3k/2}
\nonumber\\
& \quad + \frac{4}{9}\frac{1}{k^2} \int_0^{3k/2} dq qS_\calR\left( q,\frac{1}{2},\frac{1}{2} \right)
+  S_\calR\left( \frac{3}{2}k,\frac{1}{2},\frac{1}{2} \right) \, .
\end{align}
By using the consistency relation~\cite{Maldacena:2002vr} 
\begin{equation}
x_{sq}y_{sq} S_\calR\left( k,x_{sq},y_{sq} \right) = \frac{1}{4} \partial_{\log k}\log\calP_\calR \, ,
\end{equation}
we can rewrite \eqref{eq:sq-eq-enf} as
\begin{align}
\label{eq:sq-eq-enf-2}
S_\calR(k,1,1) & = \frac{61}{96}  \partial_{\log k}\log  \calP_\calR \Big|_{3k/2}
+\frac{1}{96}  \partial^2_{\log k}\log \calP_\calR \Big|_{3k/2}
\nonumber\\
& \quad + \frac{4}{9}\frac{1}{k^2} \int_0^{3k/2} dq qS_\calR\left( q,\frac{1}{2},\frac{1}{2} \right)
+  S_\calR\left( \frac{3}{2}k,\frac{1}{2},\frac{1}{2} \right) \, .
\end{align}
This relation implies that there is a minimum level of primordial non-Gaussianity controlled by the spectral index and its running as
\begin{equation}
\fnl^\text{eq} \approx 0.7 \times \partial_{\log k}\log  \calP_\calR  + 0.01 \times \partial^2_{\log k}\log \calP_\calR + \cdots ,
\end{equation}
with dots representing the contributions from other points in the $(x,y)$ plane.

Interestingly, in~\cite{Cabass:2016cgp}, it was argued that there is a non-zero minimum level of equilateral type non-Gaussianity for single field slow-roll inflation given by an amplitude of
\begin{equation}
\fnl^\text{eq} \sim 0.1 \times \partial_{\log k}\log  \calP_\calR \, .
\end{equation}
This result was derived using the transformation to local inertial coordinates focusing on how late-time observers measure the primordial bispectrum and thus the connection with our finding is not that straightforward. However, our net result is of similar spirit as that of~\cite{Cabass:2016cgp}, in saying that the lower value of primordial non-Gaussianity is controlled by the spectral index.

The shape consistency relations \eqref{eq:con-rel1} and \eqref{eq:con-rel3} are the main results of this section: since these relations are derived based on the assumption of sharp features, which in general favours higher derivative terms compared to lower ones, they represent tests of such an assumption, i.e. they provide a consistency check of sharp features in the bispectrum. This implies that if rapid time variations of the background quantities, such as the speed of sound or the slow-roll parameters, did indeed take place during inflation and features are generated, then the corresponding bispectrum evaluated for any three triplets of modes should obey \eqref{eq:con-rel1} and \eqref{eq:con-rel3}.

\subsection{Example: Starobinsky model}

\begin{figure}[h]
 \begin{center}
  \includegraphics[width=.9\textwidth]{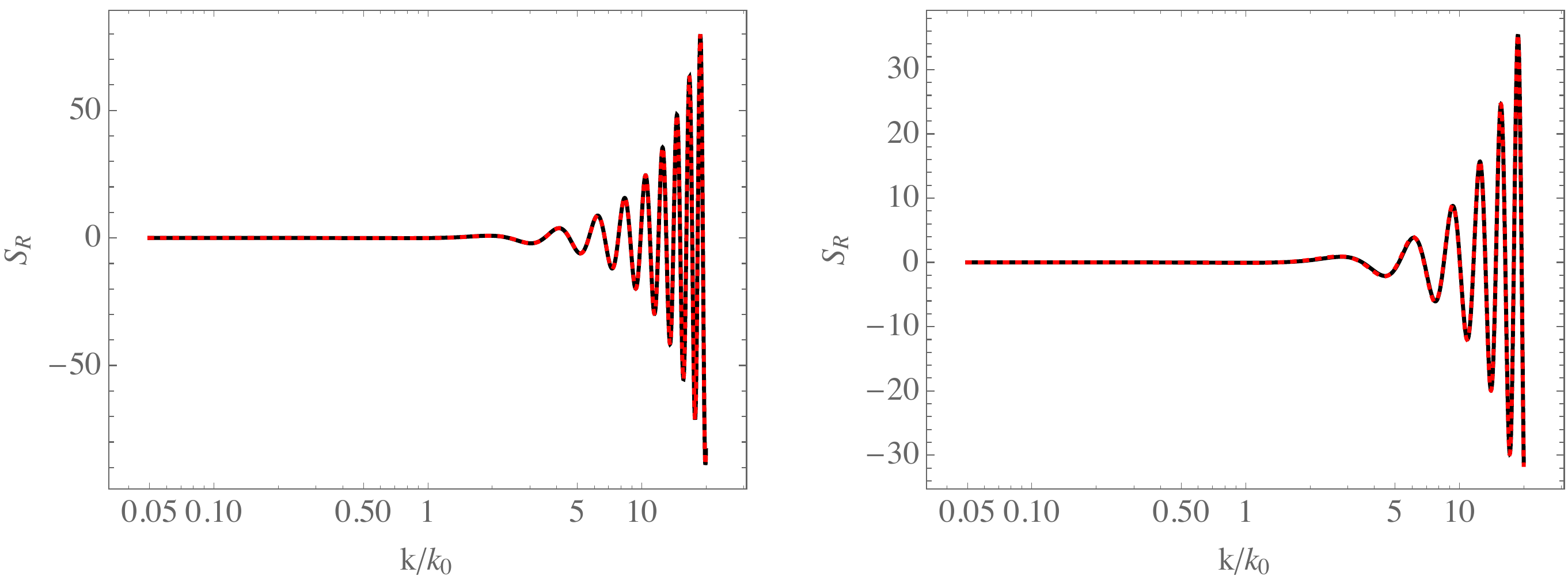}
 \end{center}
 \caption{(Left panel) equilateral ($x=y=1$) and (right panel) folded ($x=y=1/2$) shape functions for the Starobinsky model, constructed in terms of two configurations $(x_1,y_1) = (4/3,5/3)$ and $(x_2,y_2) = (3/2,7/4)$. We show the full analytic results \eqref{eq:shape-starobinsky} as solid lines and our shape consistency relation \eqref{eq:con-rel1} as dotted lines. We have set the fractional change of the slope as $\Delta{A}/A = 0.1$.}
 \label{fig:eq-folded}
\end{figure}

In order to test the validity of our shape consistency relations, we use the Starobinsky model~\cite{Starobinsky:1992ts}. It is a simple inflation model with a linear potential that experiences a sudden kink in the slope. The potential can be written as
\begin{equation}
V(\phi) = V_0 \Big\{ 1 - \left[ A + \theta(\phi-\phi_0) \Delta A \right] (\phi-\phi_0) \Big\} \, ,
\end{equation}
so that the slope changes at $\phi=\phi_0$ from $A$ to $A+\Delta{A}$ with the value of the potential $V(\phi_0)=V_0$. For simplicity, we assume that the fractional change in the slope is very small, $\Delta{A}/A\ll1$ and the de Sitter approximation holds everywhere. Then the departure from otherwise smooth slow-roll evolution is coming from $\eta$ as~\cite{Choe:2004zg}
\begin{equation}
\eta \approx -2\frac{\Delta{A}}{A} \delta(\log\tau_0-\log\tau) \, ,
\end{equation}
where $\tau_0$ is the conformal time at which $\phi=\phi_0$. Then, with $c_s= 1$, we have $c_1=-\eta$ and $c_2=\eta$ for \eqref{eq:S3}, and it is straightforward to compute the shape function $S_\calR$ as
\begin{align}
\label{eq:shape-starobinsky}
S_\calR(\kappa,x,y) = \frac{\Delta{A}}{A} & \bigg[ \frac{y}{2x} \kappa \Big\{ -\sin[(1+x+y)\kappa] + y\kappa \cos[(1+x+y)\kappa] \Big\} + \text{2 perm}
\nonumber\\
& + \frac{1+x^2+y^2}{4xy} \frac{1}{\kappa} 
\Big\{ \left[ 1-(x+y+xy)\kappa^2 \right] \sin[(1+x+y)\kappa] 
\nonumber\\
& \hspace{7em} + \left[ -(1+x+y)\kappa + xy\kappa^3 \right] \cos[(1+x+y)\kappa] \Big\} \bigg] \, ,
\end{align}
where $\kappa \equiv k/k_0 = -k\tau_0$.

\begin{figure}[h]
 \begin{center}
  \includegraphics[width=.45\textwidth]{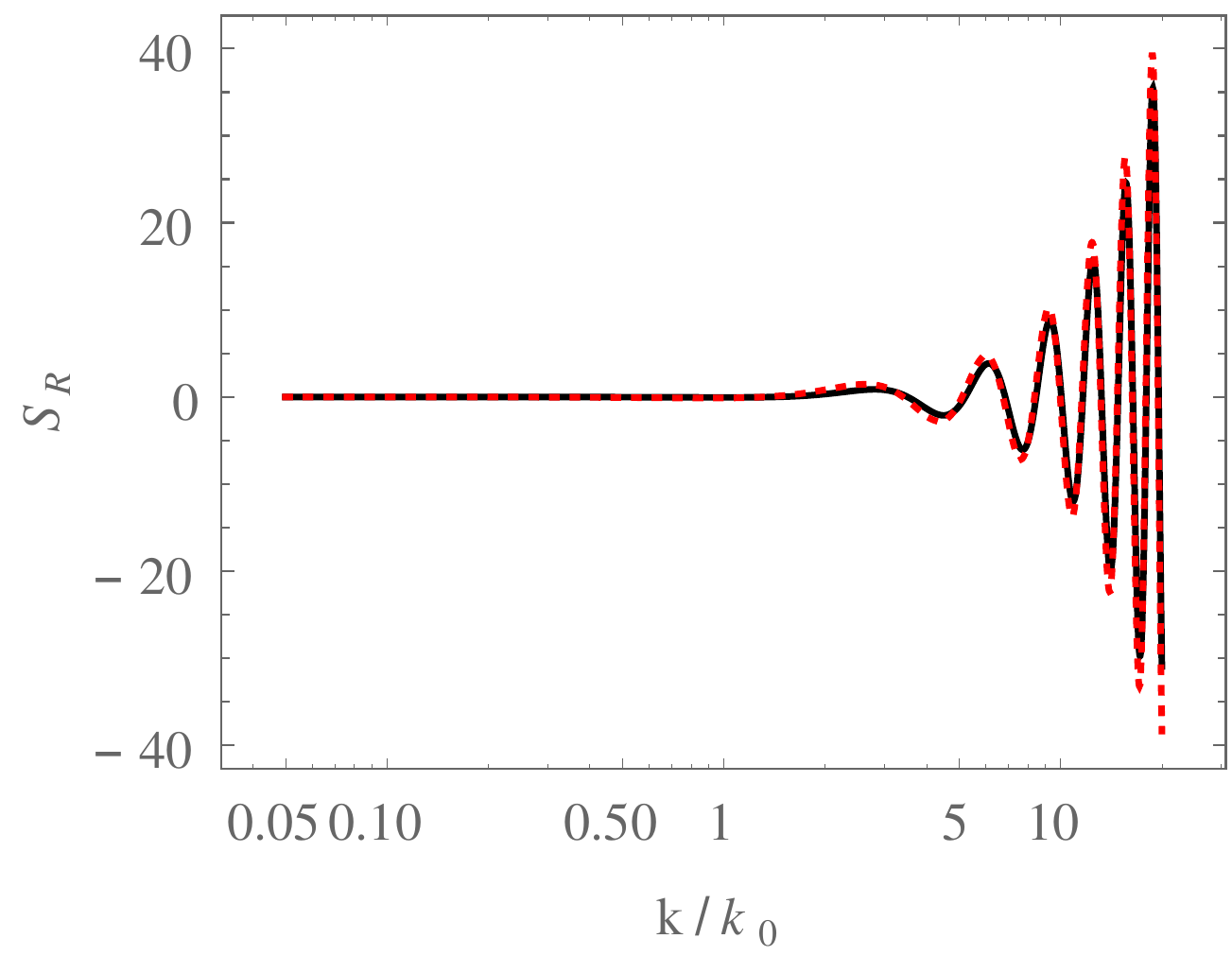}
 \end{center}
 \caption{Folded shape function constructed by those in the squeezed and equilateral configurations using \eqref{eq:con-rel3}. We show the full analytic result \eqref{eq:shape-starobinsky} as a solid line and our shape consistency relation as a dotted line. We have set $\Delta{A}/A = 0.1$.}
 \label{fig:squeezed}
\end{figure}

In Figure~\ref{fig:eq-folded}, with $\Delta{A}/A = 0.1$, we apply two configurations $(x_1,y_1) = (4/3,5/3)$ and $(x_2,y_2) = (3/2,7/4)$ to \eqref{eq:con-rel1} to reproduce other shapes. In the left and right panels, we reproduce respectively the equilateral ($x=y=1$) and folded ($x=y=1/2$) configurations by \eqref{eq:con-rel1} and compare with the analytic result \eqref{eq:shape-starobinsky}. As can be seen, our shape consistency relation \eqref{eq:con-rel1} shows excellent agreement. In Figure~\ref{fig:squeezed}, for which we set $\Delta{A}/A = 0.1$, we show the results including the squeezed shape. We use the squeezed and equilateral shapes to reproduce the folded one by using \eqref{eq:con-rel3}. Again, our shape consistency relation \eqref{eq:con-rel3} agrees nicely with the full analytic result.

\section{Templates for featured bispectrum}
\label{sec:template}

In this section, motivated by the form of the shape consistency relations, we propose two generic 2-parameter templates that could capture the featured bispectrum stemming from the time variation of both the speed of sound $c_s$ and the slow-roll parameter $\epsilon$. By thinking of $S_\calR \left( \frac{1+x+y}{1+x_i+y_i}k,x_i,y_i \right)$ in \eqref{eq:con-rel1} and \eqref{eq:con-rel3} as trigonometric functions, typical of inflationary models yielding oscillatory spectra~\cite{generalizedfeature,Ade:2015ava}, we may consider \eqref{eq:con-rel1} and \eqref{eq:con-rel3} as an expansion of the bispectrum shape in a basis and focus on the component shapes. In addition, by adopting such a view and identifying $\omega(1+x_i+y_i)^{-1}\to \omega_i$, it is clear that there could be more than one frequency at play in the oscillating $S_\calR(\omega,k,\phi)$ functions. Such a multi-frequency distribution of features is supported by the current Planck data~\cite{Ade:2015ava}.

Hence, we suggest the following two modulating templates:
\begin{align}
\label{eq:template}
S_{\alpha\beta}(x,y) & = \frac{(x+y+xy)\alpha - (1+x^2+y^2)\beta}{(1+x+y)^2} \, ,
\\
\label{eq:template2}
S_{\gamma}(x,y) & =  \gamma_1 \left[ -S_1(x,y) + \frac{F_1(x,y)}{2xy} \right] + \gamma_2 \left[ S_2(x,y)  + \frac{F_2(x,y)}{2xy} \right] \, ,
\end{align}
with the $S_i$ and $F_i$ functions defined in \eqref{s-f-defs} and the numerical coefficients $\alpha$, $\beta$, $\gamma_1$ and $\gamma_2$ in \eqref{eq:ab-def} and \eqref{g-defs}. Note that although there are more than one shape functions involved in \eqref{eq:con-rel1} and \eqref{eq:con-rel3}, they are quite degenerate leading to similar results for a wide range of parameters $\alpha$, $\beta$, $\gamma_1$ and $\gamma_2$.

\begin{figure}
 \begin{center}
  \includegraphics[width=1.\textwidth]{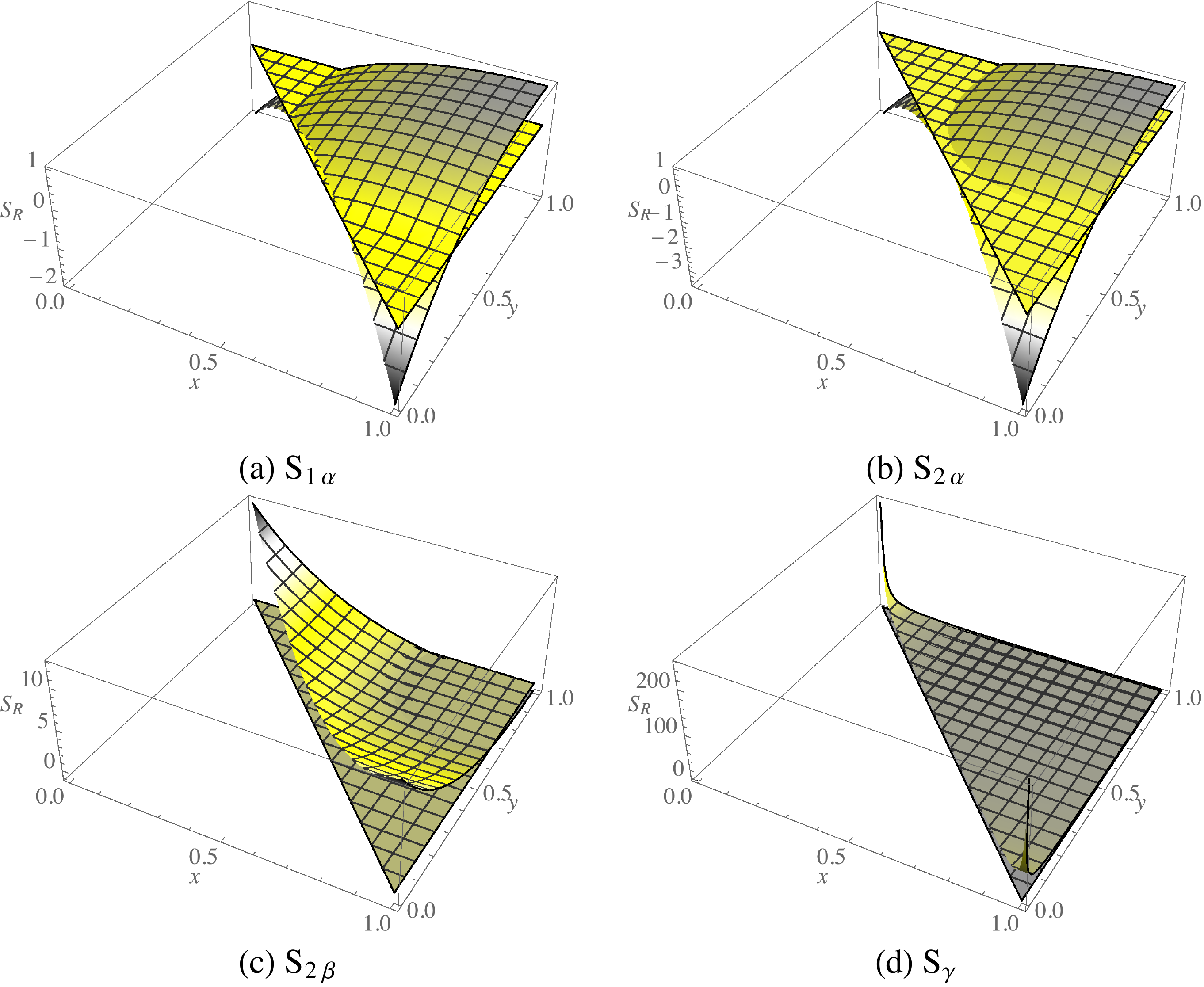}
 \end{center}
 \caption{The shapes \eqref{eq:shapes} stemming from the general templates $ S_{\alpha\beta}$ and $ S_{\gamma}$ for the representative sets of triangles given in the main text. The zero plane is also plotted. Note that the intersection with the zero plane is critical in deciding if their overlap is larger with the equilateral, orthogonal or flat templates: see Table~\ref{table:cos}.}
 \label{fig:shapes}
\end{figure}

Depending on the triangle configurations chosen as a basis, the coefficients $\alpha$, $\beta$ and $\gamma_1$, $\gamma_2$ take specific values. In order to demonstrate the generality of the proposed modulating template \eqref{eq:template}, we choose two representative sets of two triangles:
\begin{equation} 
\label{eq:triangles}
\left( \Delta_1 | \Delta_2 \right)_i = (1,2 | 2,3)_1 \quad \text{and} \quad \left( \frac32, \frac74 \Big| 2, \frac73 \right)_2 \, ,
\end{equation}
where, as in the previous section, we have specified the triangles by their lengths $\Delta_i=(x_i,y_i)$, with the remaining side being fixed to 1. These lead to the following doublets $(\alpha_1,\alpha_2)_i$ or $(\beta_1,\beta_2)_i$:
\begin{equation} 
\label{eq:doubs}
(\alpha_1,\alpha_2)_1 = (56, 44) \, , 
\quad 
(\alpha_1,\alpha_2)_2 = (41, 35) \, , 
\quad 
(\beta_1,\beta_2)_2 = (39, 37) \, .
\end{equation}
For the template \eqref{eq:template2} it is sufficient to consider the triangle $\Delta = (2,3)$, leading to 
\begin{equation} 
\label{eq:g-doubs}
(\gamma_1,\gamma_2) = \left( \frac{14}{9},\frac{22}{9} \right) \, ,
\end{equation}
since the other choices lead to similar shapes.
Plugging these values in \eqref{eq:template} and \eqref{eq:template2}, we obtain the following shapes:
\begin{equation}
\label{eq:shapes}
\begin{split}
S_{1\alpha} & = \frac{56(x+y+xy) - 44(1+x^2+y^2)}{(1+x+y)^2} \, ,
\\
S_{2\alpha} & = \frac{41(x+y+xy) - 35(1+x^2+y^2)}{(1+x+y)^2} \, ,
\\
S_{2\beta} & = \frac{-39(x+y+xy) + 37(1+x^2+y^2)}{(1+x+y)^2} \, ,
\\
S_{\gamma} & = \frac{-28(x+y+xy) + 22(1+x^2+y^2)}{18(1+x+y)^2} 
\\ 
& \quad +\frac{11 (x + y + x y) (1 + x^2 + y^2)+14 (x^2 +y^2 + x^2 y^2) }{18xy(1+x+y)} \, ,
\end{split}
\end{equation}
which are plotted in Figure~\ref{fig:shapes}. Note that these specific choices of triangles do not mean loss of generality. There should be an infinite number of triangles which lead to degenerate results. Here we only intend to demonstrate the realisation of the standard templates with specific examples.

In order to quantitatively compare these shapes with the equilateral~\cite{Creminelli:2005hu}, enfolded~\cite{Meerburg:2009ys}, orthogonal~\cite{Senatore:2009gt} and local templates~\cite{Gangui:1993tt}, which are given respectively by
\begin{equation}
\begin{split}
S_\text{eq} & = -\frac{k_3^2}{k_1k_2} - \text{(2 perm)} + \frac{k_2}{k_1} + \text{(5 perm)} - 2
\\
& = -\frac{1+x^3+y^3}{xy} + \frac{x+y+x^2+y^2+x^2y+xy^2}{xy} - 2 \, ,
\\
S_\text{enf} & = \frac{1+x^3+y^3}{xy} - \frac{x+y+x^2+y^2+x^2y+xy^2}{xy} + 3 \, ,
\\
S_\text{ortho} & = -3\frac{1+x^3+y^3}{xy} + 3\frac{x+y+x^2+y^2+x^2y+xy^2}{xy} - 8 \, ,
 \\
S_\text{loc} & = \frac{1+x^3+y^3}{3xy}  \, ,
\end{split}
\end{equation}
we employ the cosine estimator~\cite{cosine}. We define the weighted inner product of two shapes $S_1(x,y)$ and $S_2(x,y)$ as
\begin{equation}
S_1 \star S_2 \equiv \int dx dy \frac{S_1(x,y)S_2(x,y)}{1+x+y} \, ,
\end{equation}
with the integration region running over $0<x<1$ and $1-x<y<1$, and from this we define the cosine between these shapes as
\begin{equation}
\label{eq:cosine}
\cos\left(S_1,S_2\right) \equiv \frac{S_1 \star S_2}{\sqrt{S_1 \star S_1}\sqrt{S_2 \star S_2}} \, .
\end{equation}
The cosines between the standard templates and the shapes \eqref{eq:shapes} are presented in Table~\ref{table:cos}, where we can see that these templates do indeed overlap with the standard ones. We find that the orthogonal template, predicted by the effective field theory of inflation appears as a possible candidate modulating the oscillatory bispectrum. For example, while the shape $S_{2\beta}$ can be efficiently probed by the enfolded template, the shape $S_{2\alpha}$ is closer to the orthogonal one and $S_{\gamma}$ has a clear local form.

\begin{table}[h!]
 \centering
 \begin{tabular}{|l||c|c|c|c|}
 \hline
  $\cos(S_i,S_j)$ & $S_{\text{eq}}$  & $S_{\text{enf}}$ & $S_{\text{ortho}}$  & $S_{\text{loc}}$  
  \\
  \hline\hline
  $S_{1\alpha}$   & ${\bf 0.77}$ & $0.02$ & $0.52$ & $-0.09$
  \\
  \hline
  $S_{2\alpha}$   & $0.33$ & $-0.49$ & ${\bf 0.80}$ & $-0.37$
  \\
  \hline
  $S_{2\beta}$   & $0.28$ & ${\bf 0.88}$ & ${\bf -0.81}$ & $0.55$
  \\
  \hline
  $S_{\gamma}$   & $0.39$ & $ 0.56$ & $-0.36$ & ${\bf 0.99}$ 
  \\
  \hline
 \end{tabular}
 \caption{The values of the cosine estimator $\cos(S_i,S_j)$ defined in \eqref{eq:cosine} between the standard templates used by Planck and the shapes \eqref{eq:shapes}.}
 \label{table:cos}
\end{table}

Thus, taking into account the oscillating trigonometric functions, we may write general templates for the featured bispectrum as
\begin{equation}
\begin{split}
S_\text{res-$\alpha\beta$}(k,x,y) & = S_{\alpha\beta}(x,y) \sin \Big\{ C\log[(1+x+y)k]+\phi \Big\} \, , 
\\
S_\text{feat-$\alpha\beta$}(k,x,y) & = S_{\alpha\beta}(x,y) \sin \Big[ (1+x+y)\omega k+\phi \Big] \, , 
\\
S_\text{res-$\gamma$}(k,x,y) & = S_{\gamma}(x,y) \sin \Big\{ C\log[(1+x+y)k]+\phi \Big\}  \, , 
\\
S_\text{feat-$\gamma$}(k,x,y) & = S_{\gamma}(x,y) \sin \Big[ (1+x+y)\omega k+\phi \Big] \, , 
\end{split}
\end{equation}
where the labels ``res" and ``feat" have been adopted from~\cite{Ade:2015ava}, corresponding to the cases of resonant~\cite{Chen:2010bka} or generalized feature models~\cite{generalizedfeature}, and the modulating shapes $S_{\alpha\beta}$ and $S_\gamma$ are given by our general templates \eqref{eq:template} and \eqref{eq:template2} respectively. Note that these templates are suggested along the lines of those used by Planck in search for features modulated by equilateral and flattened shapes. We may well add for example a damping envelope such as $\omega k/\sinh(\omega k)$, in order to capture the fact that features are localised in $k$-space.

\section{Concluding remarks}
\label{sec:conclusion}

A variety of models inspired by UV complete theories can predict deviations from standard slow-roll inflation, leading to an oscillatory behaviour of the CMB spectrum, a signature which is marginally supported by the current observational data.

Exploiting the quasi de Sitter phase during inflation and the fact that any observable scale-dependence in the data should stem from a sharp feature in the background parameters, we have derived a model-independent set of shape consistency relations for such inflationary features in the bispectrum. Specifically, we have shown that if temporary deviations from standard slow-roll did happen during inflation, then measurements of the bispectrum for any three triplets of modes with momenta $({k}_1,{k}_2,{k}_3)_i$ for $i=1,2,3$ should be related via \eqref{eq:con-rel1} and \eqref{eq:con-rel3}.

Motivated by the form of the consistency relations, we have also produced generic two-parameter templates that include the equilateral, flattened, orthogonal and local ones, albeit with non-zero overlap. This specifically implies that the orthogonal and local templates can also appear as  modulating shapes of the oscillatory bispectrum resulting from this class of models and a search for these profiles could improve constraints on features. The form of the consistency relations also allows for templates with more than one frequency at play in the oscillatory functions in accordance to the multi-frequency distribution suggested by Planck.

An interesting generalisation of our result in this direction is to keep all terms in the Lagrangian and also all time derivatives of the coefficients $c_i$, which represent collectively background inflationary parameters. Although the result is quite complicated to be used in the form of a consistency relation, it is exactly this complexity that gives us freedom to produce several possibly new bispectrum templates. Finally, let us stress that our method could be used even in the case that features are not confirmed in order to produce possible new shapes of non-Gaussianity.

\subsection*{Acknowledgements}

We thank  
Xingang Chen and Daniel Meerburg
for useful discussions and comments. 
JG is grateful to DFI, FCFM, Universidad de Chile for hospitality while this work was under progress. 
JG acknowledges the support from the Korea Ministry of Education, Science and Technology, Gyeongsangbuk-Do and Pohang City for Independent Junior Research Groups at the Asia Pacific Center for Theoretical Physics. JG is also supported in part by a TJ Park Science Fellowship of POSCO TJ Park Foundation and the Basic Science Research Program through the National Research Foundation of Korea Research Grant NRF-2016R1D1A1B03930408. 
GAP is supported by the Fondecyt project 1171811.
SS is supported by the Fondecyt 2016 Post-doctoral Grant 3160299.

\end{document}